\documentclass[12pt,twoside]{article}
\usepackage{fleqn,espcrc1}
\usepackage{graphicx}

\title{Evolution and Nucleosynthesis of Massive Stars and Related
Nuclear Uncertainties}

\author{T. Rauscher\address{Departement f\"ur Physik und Astronomie,\\
Universit\"at Basel, \\
CH-4056 Basel, Switzerland\\
E-mail: Thomas.Rauscher@unibas.ch}
}

\begin{document}

\maketitle

\begin{abstract}
\noindent
Properties of atomic nuclei important for the prediction of
astrophysical reaction rates are reviewed. In the first part, a recent
simulation of evolution and nucleosynthesis of stars between 15 and 25
$M_\odot$ is presented. This study is used to illustrate the required
nuclear input as well as to give examples of the sensitivity to certain
rates. The second part focusses on the prediction of nuclear rates in
the statistical model (Hauser-Feshbach). Some of the important
ingredients are addressed. Discussed in more detail are approaches to
predict level densities, parity distributions, and optical
$\alpha$-nucleus potentials.
\end{abstract}

\section{Introduction}

The knowledge of nuclear reactions is crucial to model the evolution of stars
and to determine the amounts of produced nuclei.
From the abundance patterns of the
elements we can learn something about how they were created and
consequently it is possible to study the conditions in supposed
astrophysical sites and, finally, to attempt to trace the origin and
history of the Universe.
An in-depth knowledge
of the sub-atomic processes is required in order to understand many
large-scale effects.

The demands of astrophysics
challenge our ability to describe and predict nuclear properties and nuclear
reactions. Due to high temperatures and densities in the stellar plasma, a large
number of unstable nuclides are involved in the nucleosynthetic processes. Some
of their radioactive decay lines can be observed by modern satellite observatories
and provide an important tool to test the hydrodynamics of stellar models by
comparison. However, this task can only be performed if the producing nuclear
reactions are sufficiently well known. In this respect, not only reactions involving
unstable targets are important but also such along the line of stability. Despite
of temperatures in the range of millions to billions of Kelvin, the respective
energies of the interacting particles are quite low by nuclear physics standards,
from thermal energies up to a few MeV. Due to the size of the reaction networks
(the nucleosynthesis calculations for massive stars, which are described below,
include about 2400 nuclides in more than 15000 reactions), theoretical predictions
will always remain important. However, the relevance of the low-energy region
poses problems to both experimental and theoretical approaches. For neutron-induced
reactions, to resolve the transition between and interplay of different reaction
mechanisms -- from direct to resonant interactions and finally to the Hauser-Feshbach
regime -- remains a challenge. Except for activation experiments, standard techniques
only measure the resonant part of the cross section. On the other hand, energy
and strength of resonances are difficult to predict with the required accuracy.
The situation is somewhat improved whenever the statistical model of nuclear
reactions (Hauser-Feshbach theory) can be applied, which uses resonance 
averages.
Fortunately, the majority of reactions in astrophysics involving the strong
interaction can be described in this approach~\cite{rtk}. When predicting
cross sections for astrophysical applications in such a way, slightly different
points are emphasized than in pure nuclear physics investigations. Since most
of the ingredients for the calculations are experimentally undetermined (in
some cases even for nuclides at or close to stability), one has to develop reliable
phenomenological or microscopic models to predict these properties with an acceptable
accuracy across the nuclear chart. Therein one has to be satisfied with a more
limited accuracy as compared to usual nuclear physics standards. Considering
the substantially larger uncertainties in many astrophysical scenarios, this
seems to be adequate. For certain selected reactions, however, the
sensitivity of the astrophysical results is so high that an accuracy of
10\% or better is necessary. Examples of this are the reactions
$^{12}$C($\alpha$,$\gamma$)$^{16}$O and ($\alpha$,n) and
($\alpha$,$\gamma$) on $^{22}$Ne, as discussed below. When studying
details in s-process branchings, cross sections have to be determined to
1\%. This is currently only possible with experiments or by a
combination of experiment and theory, e.g. when supplementing resonance
measurements with direct capture calculations.

In this review I first present new calculations of the evolution and nucleosynthesis
of massive stars, involving reaction networks of unprecedented size. A non-exhaustive
selection of important nuclear reactions is then given and their importance
is shown by the impact on the results. In the second part of the paper, the
current standard approach to predict nuclear cross sections and reaction rates
is outlined and the remaining uncertainties and challenges are illustrated by
a few examples.

\section{Nucleosynthesis in massive stars}

\subsection{The model}

Stars above $\sim10$ $M_{\odot}$ are responsible for producing most of
the
oxygen and heavier elements found in nature.  Numerous studies have been
devoted to the evolution of such
stars and their nucleosynthetic yields.
However, our knowledge of both the input
data and the physical processes affecting the evolution of these stars
has improved dramatically in recent years.  Thus, it became
worthwhile to improve on and considerably extend the previous
investigations of pre-- and post--collapse evolution and
nucleosynthesis. 
The first calculation to determine, self-consistently,
the complete synthesis of all stable nuclides in any model for a
massive star \cite{snii} is discussed here.
The calculations were performed using the
stellar evolution code KEPLER \cite{rau:WW95} with several
modifications relative to \cite{rau:WW95}
(mass loss due to stellar winds, improved
adaptive network) and updates (OPAL95 opacity tables,
neutrino loss rates).
According to the topic of this report, the focus is on giving an
outline of the updates concerning the nuclear reactions involving the
strong interaction. For further details of the calculations
and results the reader is referred to the full paper \cite{snii}.

As in \cite{rau:WW95}, two reaction networks
are used.  A small network directly
coupled to the stellar model calculation provides
nuclear energy generation, i.e.\ it is solved implicitly
for each time-step in each zone.
This smaller network
is essentially the same as in \cite{rau:WW95},
but with updated nuclear
rates as described in the following.

One of the major improvements over \cite{rau:WW95} and other stellar
models is
that, for the first time, the synthesis of all nuclides of any
appreciable abundance is followed simultaneously in an \emph{adaptive
network} of unprecedented size.  Using a library containing rate
information for 4,679
isotopes from hydrogen to astatine, the ``adaptive'' network
automatically adjusts its size to accommodate the current nuclear
flows. This saves CPU time and thus allows to perform the calculations
within
reasonable time. Because of convective coupling of zones, the same
network must be used throughout the star.

Within such a network, the impact of nuclear rates on burning and
evolution of the star can be studied fully self-consistently.
\begin{figure}[t]
\includegraphics*[width=17cm]{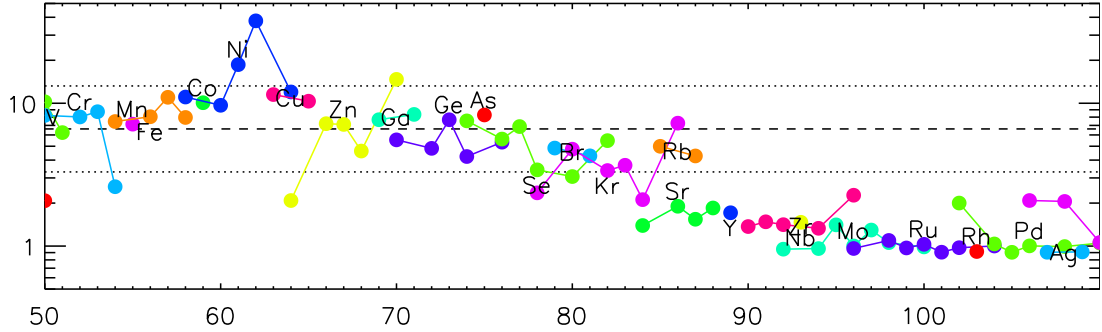}
\caption{\label{fig:s15}Postexplosive production factors (vertical axis)
versus mass number (horizontal axis)
of a population I star with 15 $M_\odot$ progenitor mass. The horizontal
axis
Comparison is relative to solar abundances.
The relative overproduction of $^{61,62}$Ni may indicate some inherent
nuclear uncertainty in the (n,$\gamma$) destruction cross section. See
text and the full paper \protect\cite{snii} for details.}
\end{figure}

\subsection{Nuclear input}
\label{nucrates}

The most extensive published library of
theoretical
reaction rates to date \cite{rt00,rt01}
was used as the backbone of the reaction rate sets. 
Details of the predictions are given in Sec.\ \ref{reaclib}.
For the network described here,
the rates based on the FRDM set were implemented.
This theoretical set was supplemented with experimental neutron capture
rates
along the line of stability \cite{bao00}.
Experimental ($\alpha$,$\gamma$) rates were
implemented for $^{70}$Ge
\cite{rau:fue96} and $^{144}$Sm \cite{rau:som98}. The derived
$\alpha$+$^{70}$Ge and $\alpha$+$^{144}$Sm potentials were also
utilized to recalculate the transfer reactions involving these
potentials.
Semi-empirical rates
were implemented for $\alpha$-capture reactions on self-conjugated
($N=Z$) nuclides \cite{rau:rtgw00}.
For the important rate
$^{12}$C($\alpha$,$\gamma$)$^{16}$O we used an updated rate
($S(300)=146$ keV barn $=1.2\times$ Ref.\ \cite{rau:buc00})
and temperature dependence
\cite{rau:buc00}. Similarly important is the
($\alpha$,n)/($\alpha$,$\gamma$) branching \cite{rau:kaepp94}
on $^{22}$Ne (see \cite{woothis}).
Table 1 defines the standard rate set.
For comparison, additional sets of experimental
and theoretical rates were used for elements below neon: Refs.\
\cite{rau:WW95,rau:HWW00}
and NACRE \cite{rau:NACRE}.
Experimental $\beta^-$, $\beta^+$, and $\alpha$-decay rates were taken
from
\cite{rau:NWC95}
and theoretical $\beta^-$ and $\beta^+$ rates from
\cite{rau:moe96}.  As a special case, a
temperature-dependent $^{180}$Ta decay
\cite{rau:end99} was implemented.  The production ratio
$^{180m}$Ta/$^{180}$Ta had to be computed offline (after production and
ejection) because the isomer was not included in the network as a
separate species. The estimate of this ratio is based on the data of
\cite{rau:end99} and a derivation given in Appendix B of \cite{snii}.

For $A\leq 40$ recent
theoretical weak rates \cite{rau:LM00} were also included.
The $\nu$-process was not followed for nuclides with $Z$ or $N$
larger than $40$ and neither was a possible r-process in high-entropy
layers close to the surface of the proto-neutron star. However, a slight
n-process could be found, due to high neutron flux generated at the base
of the He-shell \cite{snii}.

\begin{table}[tb]
\caption{Nuclear reaction rate inputs (for details see Tables 1 and 2 in
Ref.\ \protect\cite{snii}).}
\footnotesize
\begin{tabular}{ll}
\hline\hline
\multicolumn{1}{c}{Reference}&\multicolumn{1}{c}{Type}\\
\hline
Bao et al.\ (2000) \protect{\cite{bao00}} & (n,$\gamma$)\\
Buchmann (1996) \protect{\cite{rau:buc00}} &
$^{12}$C($\alpha$,$\gamma$),
modified (see Woosley, this volume)\\
Caughlan \& Fowler (1988) \protect{\cite{rau:CF88}} & light targets,
charged
projectiles\\
Fowler, Caughlan \& Zimmerman (1975) \protect{\cite{rau:fcz}} & light
targets, charged projectiles\\
F\"ul\"op et al.\ (1996) \protect{\cite{rau:fue96}} &
$^{70}$Ge($\alpha$,$\gamma$)\\
Giessen et al.\ (1994) \protect{\cite{rau:gb94}}
&$^{18}$O($\alpha$,$\gamma$)\\
Harris et al.\ (1983) \protect{\cite{rau:hfcz}} & light targets, charged
projectiles\\
Hansper et al.\ (1989) \protect{\cite{rau:ht89}} &
$^{45}$Sc($\alpha$,p),
$^{45}$Sc($\alpha$,n)\\
Iliadis et al.\ (2001) \protect{\cite{rau:id01}} & (p,$\gamma$)\\
K\"appeler et al.\ (1994) \protect{\cite{rau:kaepp94}} &
$^{22}$Ne+$\alpha$,
modified (see Woosley, this volume)\\
Kiener et al.\ (1993) \protect{\cite{rau:kl93}} &
$^{13}$N(p,$\gamma$)\\
Landr\'e et al.\ (1990) \protect{\cite{rau:la90}} &
$^{14}$N($\alpha$,p),
$^{17}$O(p,$\gamma$)\\
Mitchell et al.\ (1985) \protect{\cite{rau:mk85}} &
$^{42}$Ca+$\alpha$\\
Morton et al.\ (1992) \protect{\cite{rau:mt92}} &
$^{48}$Ti($\alpha$,p)\\
Rauscher \& Thielemann (2000,2001) \protect\cite{rt00,rt01} &
theory\\
Rauscher et al.\ (1994) \protect{\cite{rau:ra94}} &light targets,
neutron-
and $\alpha$-induced\\
Rauscher et al.\ (2000) \protect{\cite{rau:rtgw00}} & $\alpha$-capture
on
isospin symmetric targets\\
Somorjai et al.\ (1998) \protect{\cite{rau:som98}} &
$^{144}$Sm($\alpha$,$\gamma$)\\
Scott et al.\ (1992) \protect{\cite{rau:ts92}} & $^{34}$S+$\alpha$\\
Sevior et al.\ (1986) \protect{\cite{rau:sm86}} & $^{38}$Ar+$\alpha$\\
Scott et al.\ (1991) \protect{\cite{rau:sm91}} & $^{41}$K($\alpha$,p)\\
Wrean et al. (1994) \protect{\cite{rau:wb94}} & $^9$Be($\alpha$,n)\\
Wiescher \& Kettner (1982) \protect\cite{rau:wk82} &
$^{15}$O($\alpha$,p)\\
Winters \& Macklin (1988) \protect{\cite{rau:wm88}} &
$^{20}$Ne(n,$\gamma$)\\
\hline\hline
\end{tabular}
\end{table}

\subsection{A non-exhaustive selection of important reactions}

It should be noted that only a selection of a few important rates is
presented in this section. If a reaction does not show up here, this
does not imply that it is of no importance. It just means that there was
not sufficient space to discuss the reaction properly. Main reactions in
the quiescent burning phases of a star (e.g.\ $^{12}$C+$^{12}$C in
carbon burning) as well as those many neutron capture reactions
necessary for a detailed study of s-process branchings will always
remain on the list of important reactions. However, a few reactions were
chosen subjectively for discussion here.

\subsubsection{$^{12}$C($\alpha$,$\gamma$)$^{16}$O}

It is well known that the leading nuclear uncertainty afflicting modern
studies of stellar evolution and nucleosynthesis continues to be the
reaction rate of $^{12}$C($\alpha$,$\gamma$)$^{16}$O. This reaction
competes for the consumption of $\alpha$-particles
with the triple-alpha reaction during helium burning. It determines the
ratio of C to O at the onset of the subsequent burning stage, i.e.\
carbon burning. This has important implications not only for
nucleosynthesis but determines the further evolution of the star and
even the nature of the collapse and explosion.

Weaver \& Woosley \cite{WeaW93} suggested a preferred value of
170$\pm$20 keV b for the total S-factor at 300 keV, based on
nucleosynthesis arguments. Recently, those calculations were repeated
using more modern stellar models \cite{hwb03} and arrived at basically
the same conclusions as \cite{WeaW93}. For that reason, the currently
preferred choice for the absolute value and energy dependence of the
rate is the one given in Sec.\ \ref{nucrates}.

A further implication of the nucleosynthesis study is that the rate has
to be known to an accuracy of $\leq10$\%. Recent measurements 
\cite{rau:buc00,buc97,bru01,kun02,tisch}
give recommended values in the range 145 to 165 keV b but with
considerable uncertainty. Only the work of Tischhauser {\it et al.}
\cite{tisch} is beginning to approach that precision.

\subsubsection{$^{22}$Ne($\alpha$,$\gamma$)$^{26}$Mg and
$^{22}$Ne($\alpha$,n)$^{25}$Mg}

Behind $^{12}$C($\alpha$,$\gamma$)$^{16}$O, the $^{22}$Ne($\alpha$,n)
rate and the ($\alpha$,n)/($\alpha$,$\gamma$) branching at $^{22}$Ne
become the second largest nuclear uncertainties in calculating the
nucleosynthesis in massive stars.
The weak s-process component, including a large number of nuclides
produced by neutron captures up to about $A=90$, is made in massive
stars. The relative production of these nuclei compared to other
abundant species like O and to one another is sensitive to the cross
sections of neutron poisons like $^{25}$Mg and $^{16}$O \cite{woothis},
and to the rate of
$^{22}$Ne($\alpha$,n)$^{25}$Mg.

\begin{figure}[t]
\centerline{\includegraphics*[width=10cm]{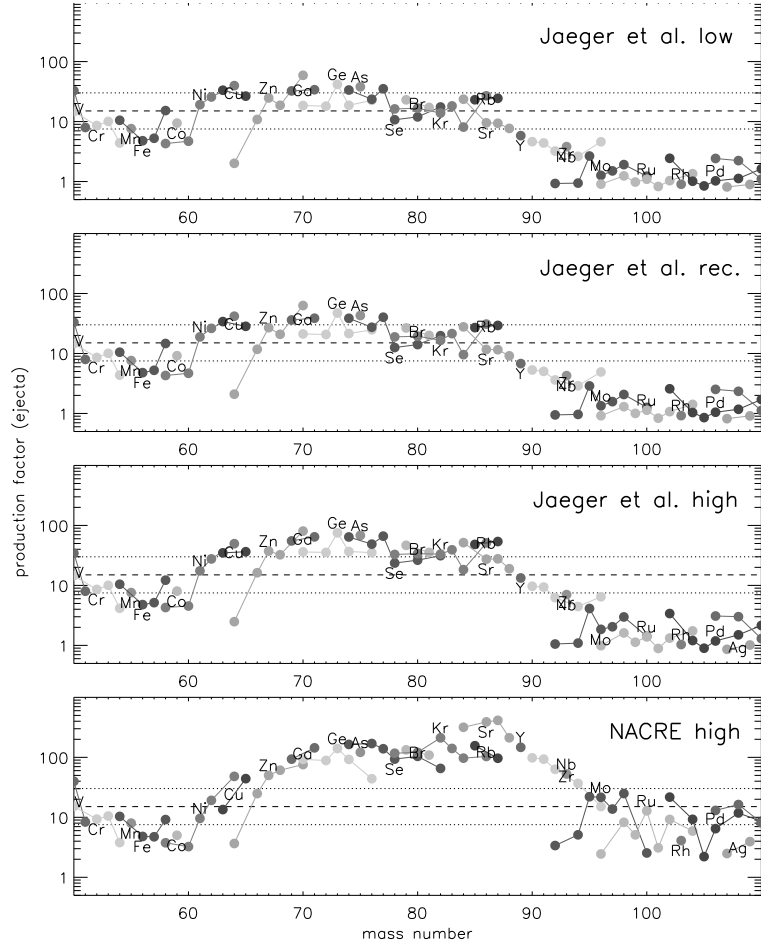}}
\caption{\label{figne22}Nucleosynthesis of the weak s-process with
varied rate of $^{22}$Ne($\alpha$,n)$^{25}$Mg in a 25 $M_\odot$ star
of initial solar metallicity. A
factor of two overproduction (dotted lines) is acceptable because lower
mass stars make less s-process and low metallicity stars make very
little.}
\end{figure}

The ($\alpha$,n) rate has always been quite uncertain \cite{rau:NACRE}.
Recent experimental work \cite{jaeger} exhibits higher accuracy but is
still uncertain enough to accommodate a factor of two uncertainty in
many important s-process products for $80\leq A\leq 90$ (see Fig.\
\ref{figne22}).

\subsubsection{The \protect\( ^{62}\protect \)Ni(n,\protect\( \gamma \protect \)) case}

This case is a good example of the difficulties encountered when trying
to predict reaction rates for final nuclei with low level densities.
For neutron-induced reactions at low energies, close to magic numbers, and far
off stability where low separation energies are encountered, a problem
emerges. In such targets, the level density is too low to allow the application
of the statistical model~\cite{rtk}. Also for other nuclides it is not straightforward
to bridge the region of thermal energies to the region of overlapping resonances
where the Hauser-Feshbach formalism can be used. Single resonances and direct
reactions become important. This is also an issue for neutron-rich nuclei in
the \( r \)-process path with low neutron-separation energies.

With the reaction $^{62}$Ni(n,$\gamma$)$^{63}$Ni it was previously
attempted to extrapolate thermal data to s-process energies of up to a
few hundred keV. 
Two compilations give disagreeing 30 keV cross sections~\cite{bao,bao00},
based on the same thermal data. Both extrapolations assume s-wave behavior of
a direct capture component. The more recent one includes a sub-threshold resonance
contributing to the thermal cross section.

\begin{figure}[t]
{\par\centering \includegraphics*[width=12cm]{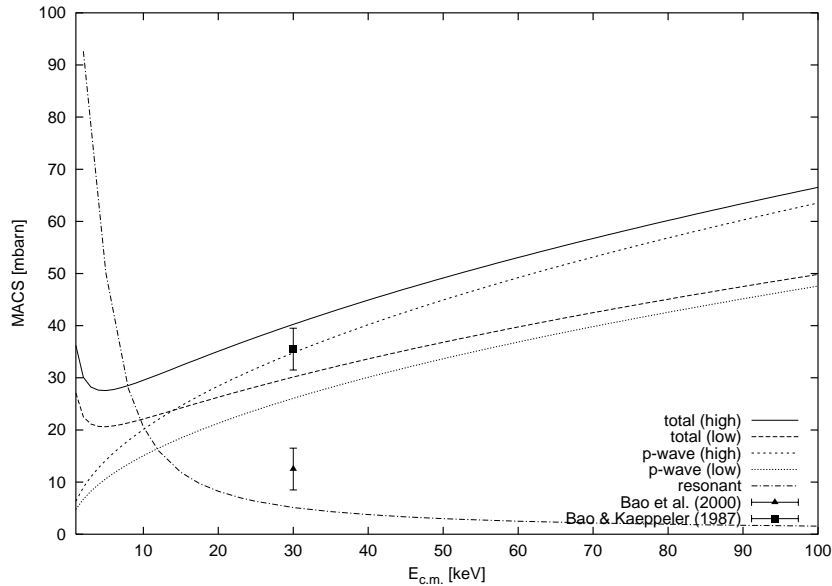}\par}

\caption{Direct neutron capture Maxwellian averaged cross section of \protect\( ^{62}\protect \)Ni.
The final value is given by adding the resonant contribution to the {}``total{}''
direct term. Upper and lower limits on the direct components are from experimental
errors on the input, \textit{i.e.} in the thermal scattering length and the
spectroscopic factors.\label{figguber}}
\end{figure}

A calculation of the direct capture component using DWBA found a considerable
p-wave contribution which enhances the cross section at 30 keV~\cite{guber}.
Thus, even when including the subthreshold resonance, the 30 keV value is coincidentally
similar to the value in the older compilation (Fig.\ \ref{figguber}). However, also the general
energy dependence of the cross section is altered. Resonances were also included
but they only contribute less than 15\%. The enhanced cross section has an important
impact on \( s \)-processing in massive stars. The previously seen overproduction
of \( ^{62} \)Ni in stellar models (see Fig.\ \ref{fig:s15})
can be cured when using the enhanced rate
because of increased destruction of this nucleus with the larger neutron capture
rate~\cite{snii,guber}. An experimental verification of this result
would be desireable.

Further neutron capture reactions on Ni isotopes might also be suspect
but they do not show the additional complication of a sub-threshold
resonance.

\section{Statistical model calculations}
\label{reaclib}

\subsection{A reaction rate library}

A recently published large-scale reaction rate library includes neutron-, proton-,
and \( \alpha  \)-induced reactions on all target nuclei from Ne up to Bi from
proton-dripline to neutron-dripline~\cite{rt00,rt01}. Due to the fact that
many very short-lived nuclides can be produced in astrophysical sites, it is
necessary to provide cross sections and rates for about 4600 targets and 32000
reactions. These numbers show that theory will always play a major role in providing
cross sections, despite the potential of future Rare Isotope Accelerators.
This rate library has
already been adopted as a standard for calculating
nucleosynthesis in stellar evolution
and in type II supernovae~\cite{snii,hegthis}.
Fits to the astrophysical reaction rates -- ready for direct
astrophysical
application -- as well as tables of cross sections, reaction rates, and
nuclear inputs for all possible reactions with light projectiles
can also be downloaded from {\it http://nucastro.org/reaclib.html} .

\subsection{Global calculations}

The calculations for the above library
were performed with the Hauser-Feshbach code NON-SMOKER~\cite{nonsm}
which
is especially tuned to such large-scale predictions.
The theoretical determination
of nuclear properties in this code is based on microscopic or
macroscopic-microscopic approaches suitable for an application far off
stability. As mentioned before, due to the large number of unstable
nuclei to be implemented in nuclear reaction networks for astrophysics,
nuclear properties needed for the calculation of reaction rates have to
be predicted. Necessarily, there is a trade-off with accuracy because
one cannot expect to predict those properties with similar accuracy as
measured data. However, whenever measured properties can be used, this
will, of course, lead to an improvement in the calculated rates.
Unfortunately, this is not possible for the vast majority of reactions
off the line of stability (and sometimes not even at stability).

Occasionally, a misconception about the best models for nuclear
properties seems to arise. The question is not merely one of microscopic
versus phenomenological approaches, it should rather be ``{\it what can we
learn from the different models and how do we implement the findings in
a simple way, suited for large-scale computations?}''. On one hand,
first-principles microscopic models can be more satisfying from a
philosophical point of view. Such a model will of course be most
preferrable as soon as the perfect one has been identified. Unfortunately, the
current state-of-art sees competition of many different microscopic
approaches, each with its own advantages and drawbacks, but we are still
far from a complete, unified picture. Thus, one has to be careful when
using the term ``microscopic'' in the sense of a general remedy. 

On the
other hand, the problem inherent in all microscopic approches is that
they are CPU expensive. Although further progress regarding CPU speed and
perhaps algorithmical effectiveness can be expected, this will hamper
large-scale studies for some time to come. A more effective approach is
to {\it understand} the important effects found in microscopic calculations
or based on {\it fundamental properties} of nuclei
and to derive a simple, even if phenomenological or parametrized, way to
implement them in the reaction rate calculations. A good example for
using microscopic (in this case, shell model Monte Carlo) calculations 
in this way is the treatment of the parity distribution in nuclei at low
excitation energies, as discussed in Sec.\ \ref{secratio}. An example of
using fundamental properties of nuclei is the approach to calculate
nuclear level densities (Sec.\ \ref{seclevden}).

In the NON-SMOKER code it was tried to implement reliable predictions of
nuclear properties following the above philosophy.
For instance, the nucleon optical potentials are calculated in a fast
microscopic model \cite{jeu}. 
The photon transmission coefficients are derived using a global
parametrization of GDR properties \cite{thiarn},
based on the hydrodynamic droplet
model and including an energy-dependent GDR width \cite{rt00}.
Further included are width fluctuation corrections and isospin effects
\cite{nonsm}.
The nuclear level density will be discussed in the following section.
Details of all nuclear properties used are given elsewhere
\cite{rt00,nonsm}.

\subsection{Nuclear Level Density}

\subsubsection{A global description}
\label{seclevden}

The nuclear level density is an important ingredient in the prediction
of nuclear reaction rates in astrophysics. The applicability of the
statistical model of nuclear reactions (Hauser-Feshbach formalism) can
be derived from the average level spacing and lower limits of energy and
temperature for the
application to calculate cross sections and reaction rates,
respectively, can be given \cite{rtk}.

The current version of the NON-SMOKER code uses a level density
description based on {\it fundamental properties} of nuclei, as introduced
above. The macroscopic properties can be derived from the Fermi-gas
formalism \cite{bohr}. Obviously, these will hold as long as the
nucleons can be described by a Fermi-gas, regardless of whether they are
close to or far from stability. Pairing is usually accounted for by
introducing a backshift, leading to the well-known expressions of the
shifted Fermi-gas \cite{bethe36}:
\begin{equation}
\rho (U,J,\pi )={\mathcal{F}}(U,J,\pi)\rho (U)\quad ,
\end{equation}
 with
\begin{eqnarray}
\rho (U)\propto {1\over \sigma a^{1/4}}{\exp
(2\sqrt{a
U})\over U^{5/4}}
\\
\sigma ^{2}={\Theta _{\mathrm{rigid}}\over \hbar ^{2}}\sqrt{U\over a}\,
\, ,\qquad
U=E-\delta \quad .\nonumber
\end{eqnarray}
 The spin and parity dependence \( {\mathcal{F}} \) is determined by the spin
cut-off
parameter \( \sigma  \). Thus, the level density is dependent on only
two parameters:
the level density parameter \( a \) and the backshift \( \delta  \),
which
determines the energy of the first excited state. The divergence for
$E=\delta$ can be avoided by either introducing an additional term
depending on the nuclear temperature \cite{lang54} or by matching it to
the constant temperature formula at low energies \cite{rtk}.

Within this framework, the quality of level density predictions depends
on the
reliability of systematic estimates of \( a \) and \( \delta  \). 
All current (microscopic) calculations prove that the backshifted Fermi-gas can
account for the nuclear level density as long as $a$ and $\delta$ are
chosen properly, e.g.\ Shell Model Monte Carlo \cite{alhassid,dean},
combinatorial approaches \cite{paar}, and recurrence relations for exact
level densities \cite{VanI}.

Thus, while the 
overall shape of the energy
dependence is given by the globally valid Fermi-gas, the microscopic
corrections enter via $\delta$ and mainly via $a$:
\begin{equation}
a(U,Z,N)=\tilde{a}(A)\left[ 1+C(Z,N){f(U)\over U}\right] \quad ,
\end{equation}
 where
\begin{equation}
\tilde{a}(A)=\alpha A+\beta A^{2/3}
\end{equation}
 and
\begin{equation}
f(U)=1-\exp (-\gamma U)\quad .
\end{equation}
Therefore, the parameter $a$ contains all effects {\it beyond} a
spherical droplet.
Here, the microscopic correction $C(Z,N)$ is thermally damped away at
high excitation energies as suggested by \cite{igna75,igna78} after
inspection of hydrodynamic models. Three free parameters $\alpha$,
$\beta$, $\gamma$ have to be
fitted to nuclear data. The strength of the above approach
lies in the fact that these {\it microscopic properties} can easily be
extracted from other (microscopic or semi-microscopic) calculations
which are available across the whole nuclear chart: nuclear mass models.
Many different inputs can easily be implemented and compared. This is
explained in detail in \cite{rtk}.

\begin{figure}[t]
\includegraphics*[width=6cm,angle=90]{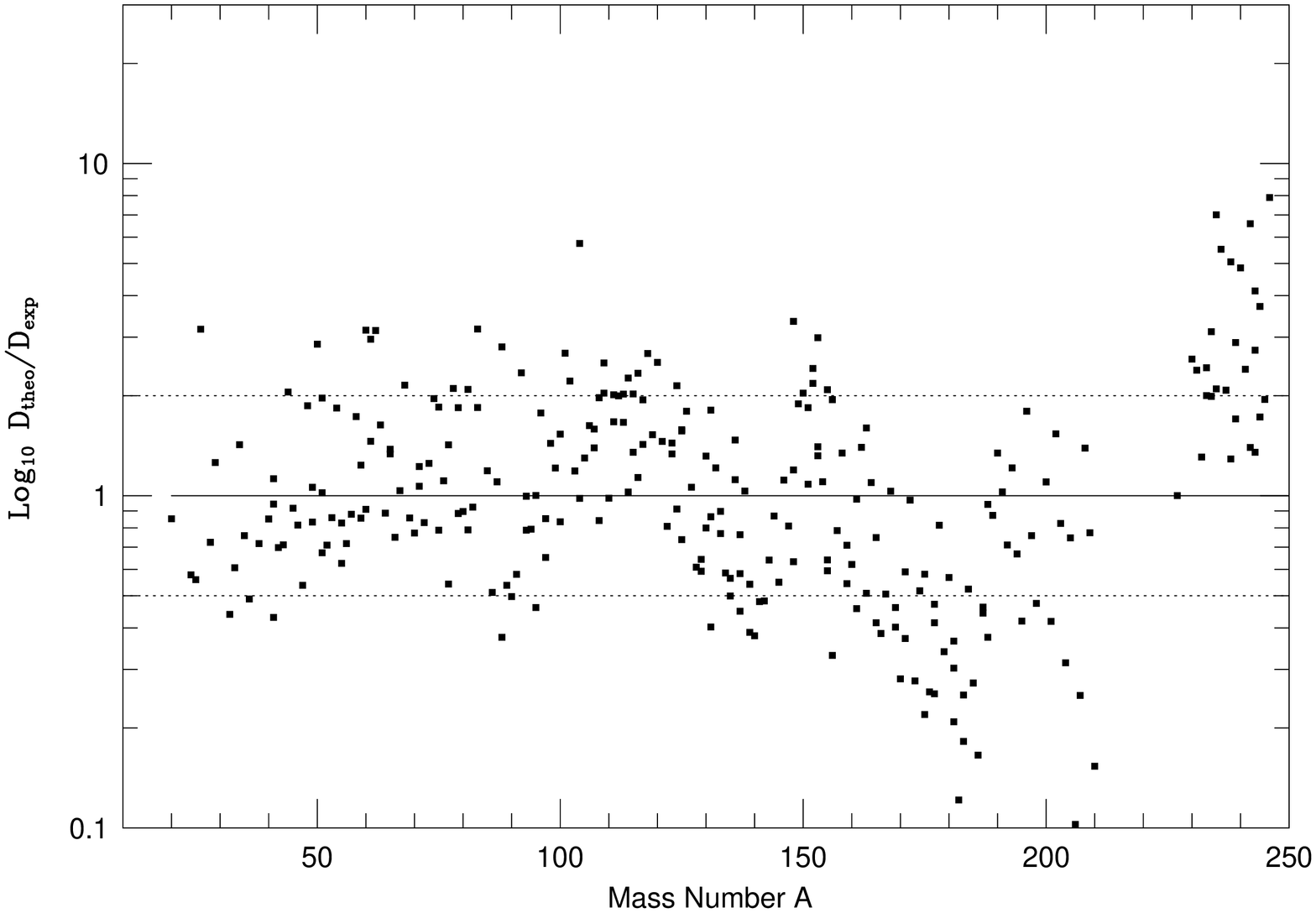}
\includegraphics*[width=6cm,angle=90]{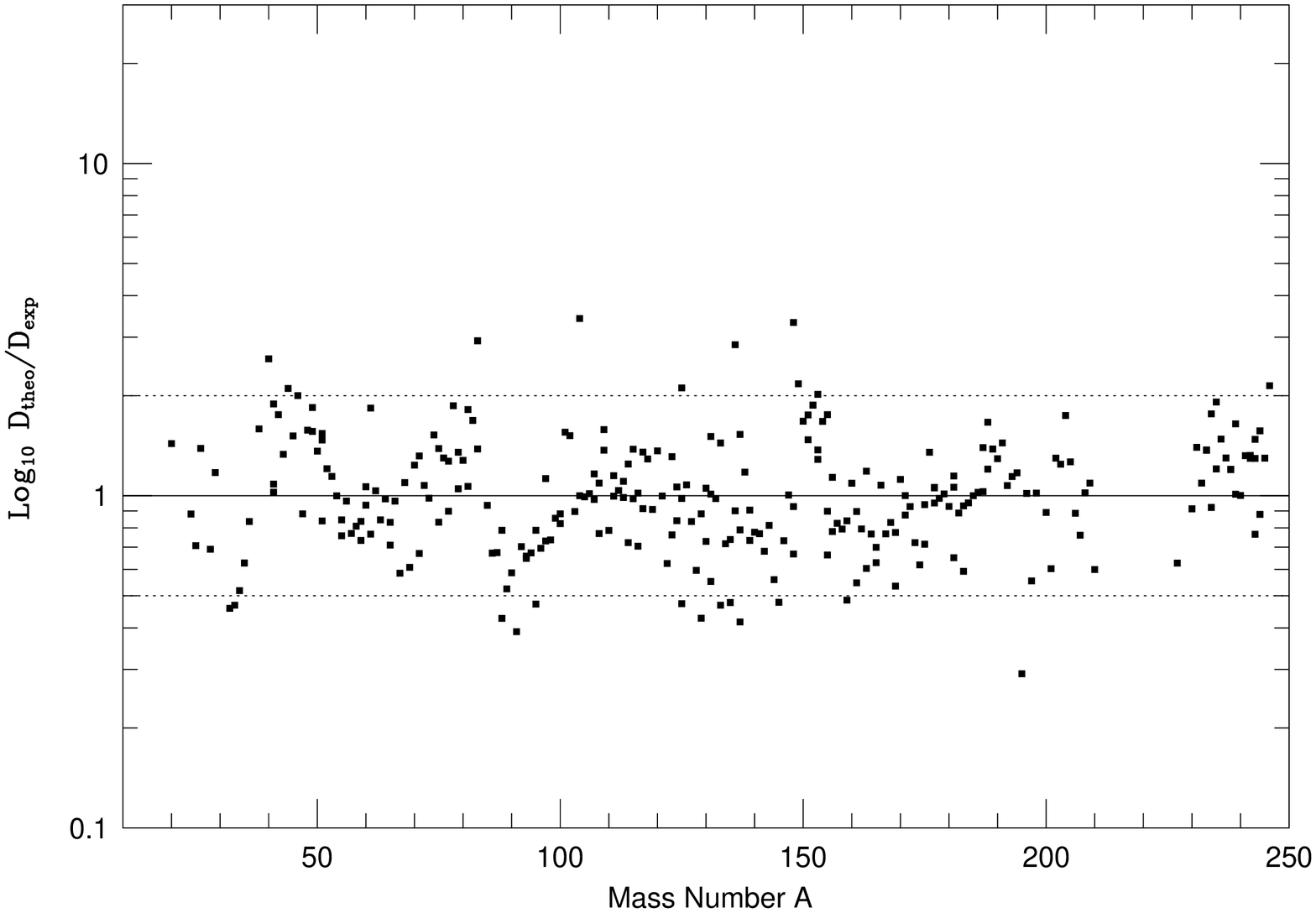}
\caption{\label{fig:levden1}Ratio of predicted to experimental level
densities at the neutron separation energy with microscopic inputs taken
from \cite{hilf} (left) and from \cite{frdm} (right). The deficiencies
of the underlying mass formula are propagated to the level density
result. For instance, the well-known problems of \cite{hilf} at higher
mass numbers can be seen clearly. Using input from \cite{frdm} improves
the situation drastically. However, shell closures are still overpronounced as
can be seen from the emerging pattern in the level density ratios.}
\end{figure}

Obviously, the
resulting level density inherits the deficiencies of the underlying
model concerning the microscopic properties. This is illustrated by
Fig.\ \ref{fig:levden1} for two cases. Nevertheless, for the
majority of nuclei deviations of less than a factor of two can be
obtained when using, for instance the FRDM input \cite{frdm}, 
which translates into a much 
lower uncertainty (about 30\%) in the reaction rates. It has to be
emphasized that the accuracy of this purely theoretical approach is
still unsurpassed, despite recent attempts for improvement with other
approaches, e.g.\ \cite{gor96,demlevden} (it should be noted that the
average deviation quoted in \cite{demlevden} includes results which were
renormalized to experiment and therefore is not a measure of predictive
power).

For all details of the level density description and further
implications, please refer to the
extended paper \cite{rtk}.

\subsubsection{Implementing parity-dependence}
\label{secratio}

So far, all theoretical,
global calculations of astrophysical rates assume an equal
distribution of the state parities at all energies. It is obvious that
this assumption is not valid at low excitation energies of a
nucleus. However, a globally applicable recipe was lacking. We combine
a formula for the energy-dependent parity distribution \cite{alha} with the
microscopic-macroscopic nuclear level density described above
\cite{darkonic,darkocgs}. The formula
reproduces well the transition from low excitation energies where a
single parity dominates to high excitations where the two densities
are equal. It was tested against Monte Carlo shell model
calculations.

Alhassid {\it et al.} \cite{alha} have introduced a simple model for the 
partition function ratio
$Z_-/Z_+$ for nuclei in the iron region using the complete {\it pf +
  g$_{9/2}$} shell. This model was combined with the shifted Fermi-gas
approach to derive parity-dependent level densities.
For details, see \cite{darkonic,darkocgs}.

\begin{figure}[t!]
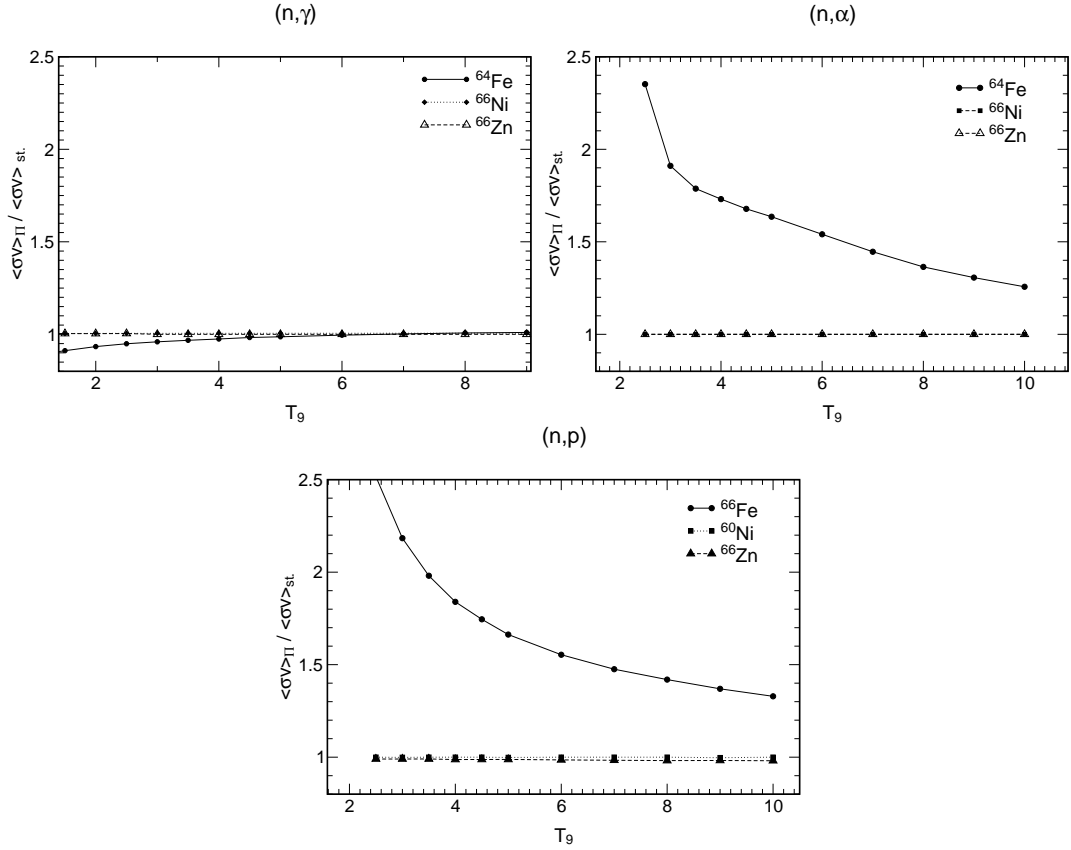

\begin{center}
\includegraphics*[width=7cm]{prag_ngamma.eps}
\includegraphics*[width=7cm]{prag_nalpha.eps}
\centerline{\includegraphics*[width=7cm]{prag_np.eps}}
\end{center}
\caption{\label{figrateratio}Comparison of the parity dependent reaction rates to the
  standard \protect\cite{rt01}, which assumes an equal distribution of
  odd and even parity states. The {\it final} nucleus is specified for
each reaction.}
\end{figure}

The parity-dependent level density was used to calculate
astrophysical reaction rates involving the three nuclides 
$^{64}$Fe, $^{66}$Ni, $^{68}$Zn
in the global Hauser-Feshbach model
NON-SMOKER. A comparison to the standard values is shown in Fig.\
\ref{figrateratio}.
The impact on the rates involving the Ni and
Zn nuclei is small and negligible compared to the remaining
uncertainties in the global HF model. This is due to the fact that a
sufficiently large number of excited states is known
experimentally. Up to 20 experimental states are considered in the
standard calculation and only above the last known state, the
theoretical level density is in effect. However, the case is different
for reactions involving $^{64}$Fe. No information on experimental
states is known here and therefore the full impact of the parity
dependence can be seen. In the (n,$\gamma$) case $20\%$ difference
are found. Much larger differences are seen in the reactions involving
$^{64}$Fe in the final particle channel. Because of lack of negative
parities at low excitation energies, the particle emission channel
becomes strongly enhanced in all such reactions with low or negative
Q values. The (n,$\gamma$) channel show lower sensitivity because the
total transmission coefficient includes more transitions to states at
higher excitation energy where the parity ratio is already close to
unity. This is not true for the particle (exit) channels where
preferrably states at low excitation energy are populated, due to the
reaction energetics.

The case for $^{64}$Fe shows that a large effect of the parity
dependence can be expected far from stability where no experimental
information on excited states is available and that it is extremely
important to include such a modified level density. The current approach
is valid only for even-even nuclei in the {\it pf + g$_{9/2}$}
shell. Work is in progress to extend this description to be able to
calculate the parity distribution for a large number of nuclei far
from stability on the proton-rich as well as neutron-rich side.

\subsection{Optical \protect\( \alpha \protect \)-nucleus potentials}

There have only been few attempts to derive global optical potentials for \( \alpha  \)-projectiles~\cite{raunic}
and most of them are only valid at \( \alpha  \)-energies larger than 30 MeV.
Due to the high Coulomb barrier and nuclear structure effects defining the imaginary
part of the potential it is difficult to obtain a global potential at astrophysical
energies. Elastic \( \alpha  \)-scattering data can constrain the real part
of the potential~\cite{mcf,mohr97} and detailed analysis can also improve
on the imaginary part~\cite{mohr00,molyb}, describing the absorption into
other channels than the elastic scattering, i.e.\ the Hauser-Feshbach channel.
Due to the scarcity of data for intermediate and heavy nuclei, attempts to improve
on the potential are mostly concentrating on single reactions~\cite{rau:som98,molyb}.
More global approaches suffer from the lack of data to confine their parameters~\cite{raunic,gornic}.

\begin{figure}[t]
{\par\centering \includegraphics*[width=13cm]{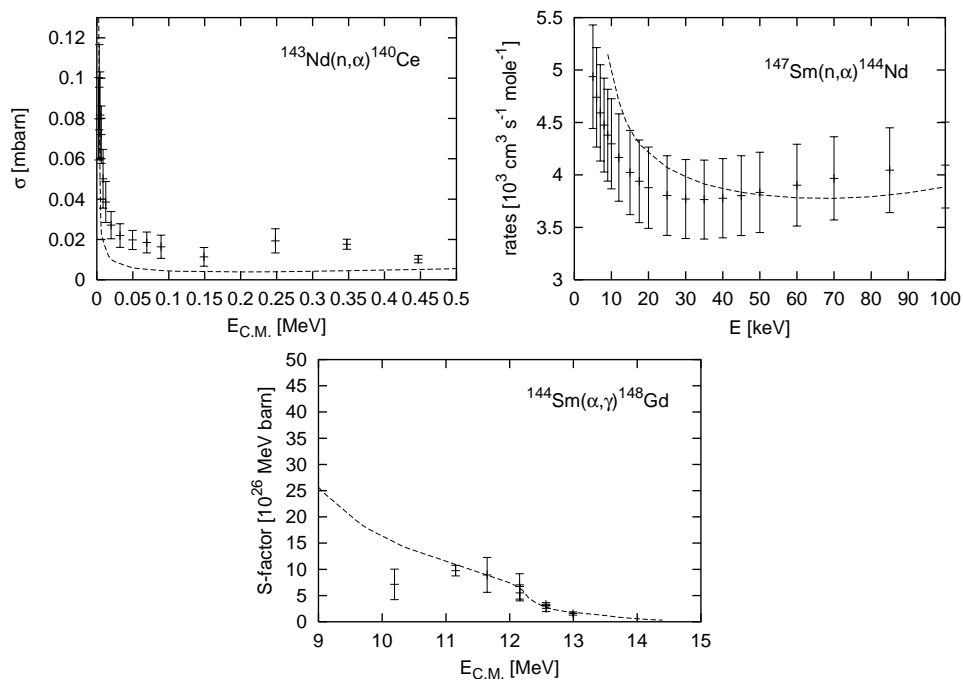}\par}

\caption{\label{figcarla1}Cross sections, reaction rates, and S-factors from a simultaneous \protect\( \chi ^{2}\protect \)
fit of the Fermi-type energy-dependent \protect\( \alpha \protect \)+nucleus
optical potentials of three reactions (see text). The dashed lines are the statistical
model calculation. The errors on the \protect\( ^{147}\protect \)Sm(n,\protect\( \alpha \protect \))
rates were assumed to be 10\%.}
\end{figure}

\begin{figure}[t]
{\par\centering \includegraphics*[width=13cm]{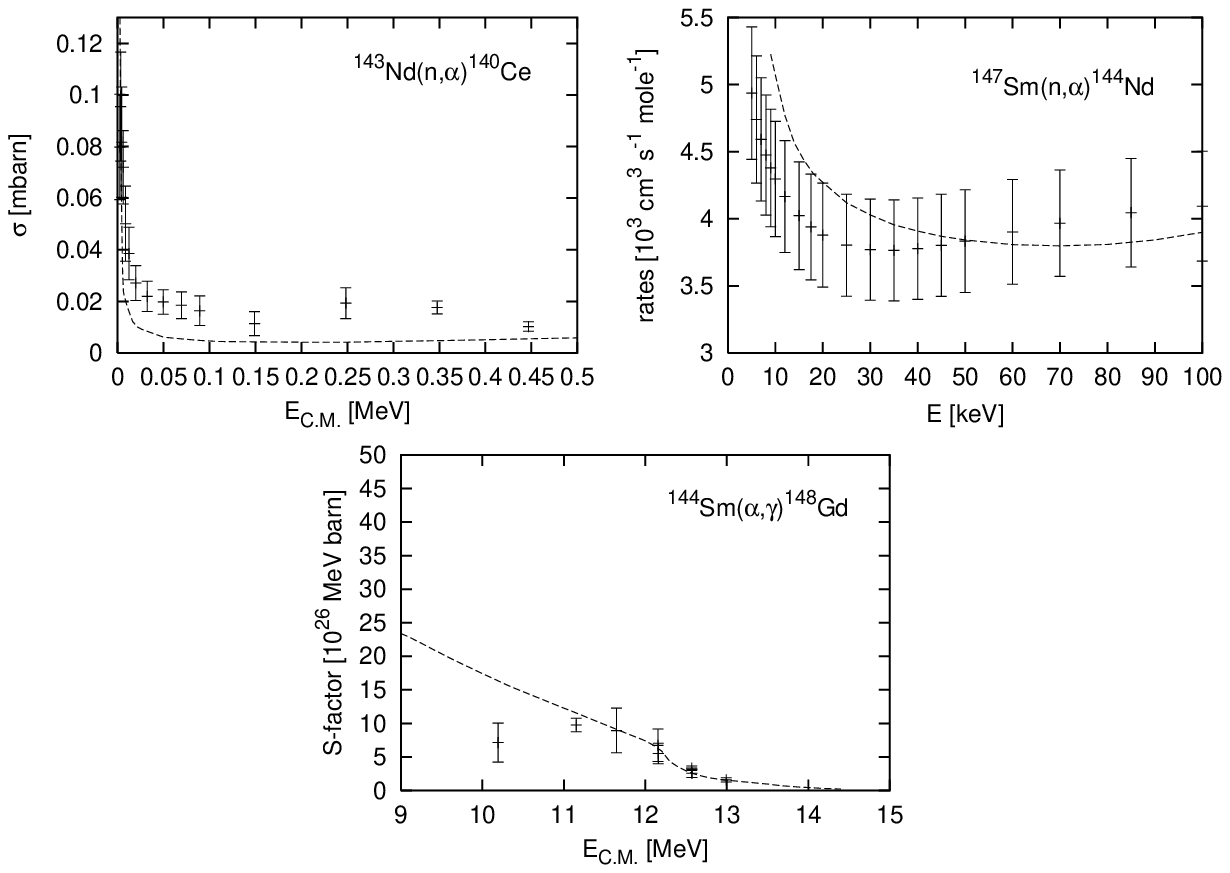}\par}

\caption{\label{figcarla2}Cross sections, reaction rates, and S-factors from a simultaneous \protect\( \chi ^{2}\protect \)
fit of the Brown-Rho energy-dependent \protect\( \alpha \protect \)+nucleus
optical potentials of three reactions (see text). The dashed lines are the statistical
model calculation. The errors on the \protect\( ^{147}\protect \)Sm(n,\protect\( \alpha \protect \))
rates were assumed to be 10\%.}
\end{figure}

Recently, it was tried \cite{froh} to find a potential for the \( A\simeq 140 \) mass region by
simultaneously fitting data for \( ^{143} \)Nd(n,\( \alpha  \))\( ^{140} \)Ce~\cite{koehpriv},
\( ^{147} \)Sm(n,\( \alpha  \))\( ^{144} \)Nd~\cite{koeh}, and \( ^{144} \)Sm(\( \alpha  \),\( \gamma  \))\( ^{148} \)Gd~\cite{rau:som98}.
The optical potential is parametrized as \begin{equation}
V(r,E)=-\frac{V_{0}}{1+\exp \left( \frac{r-r_{r}A^{1/3}}{a_{r}}\right) }-i\frac{W(E)}{1+\exp \left( \frac{r-r_{V}A^{1/3}}{a_{V}}\right) }\quad .
\end{equation}
 Different parameters for the potential geometry and the energy dependence of
the depth of the imaginary part were explored~\cite{froh,froh1}. No
significant differences were found between using a Brown-Rho shape~\cite{mohr97} \( W(E)=W_{0}((E-E_{0})^{2})/((E-E_{0})^{2}+\Delta ^{2}) \)
or a Fermi-type shape~\cite{rau:som98} \( W(E)=W_{0}/(1+\exp ((E^{*}-E)/a^{*})) \)
of the energy dependence. For the latter, the parameters \( E^{*}=18.74 \)
MeV, \( a^{*}=2.1 \) MeV were found, with all other parameters as in the previous paper~\cite{rau:som98}.
The Brown-Rho best fit was obtained with \( E_{0}=6.35 \) MeV and \( \Delta =28.4 \)
MeV, with the same fixed parameters \( V_{0}=162 \) MeV, \( r_{r}=1.27 \)
fm, \( a_{r}=0.48 \) fm, \( W_{0}=19 \) MeV, \( r_{V}=1.57 \) fm, \( a_{V}=0.6 \)
fm. The results from the simultaneous fit of three reactions are shown in 
Figs.\ \ref{figcarla1} and \ref{figcarla2}. 

\begin{figure}[t]
{\par\centering \includegraphics*[width=13cm]{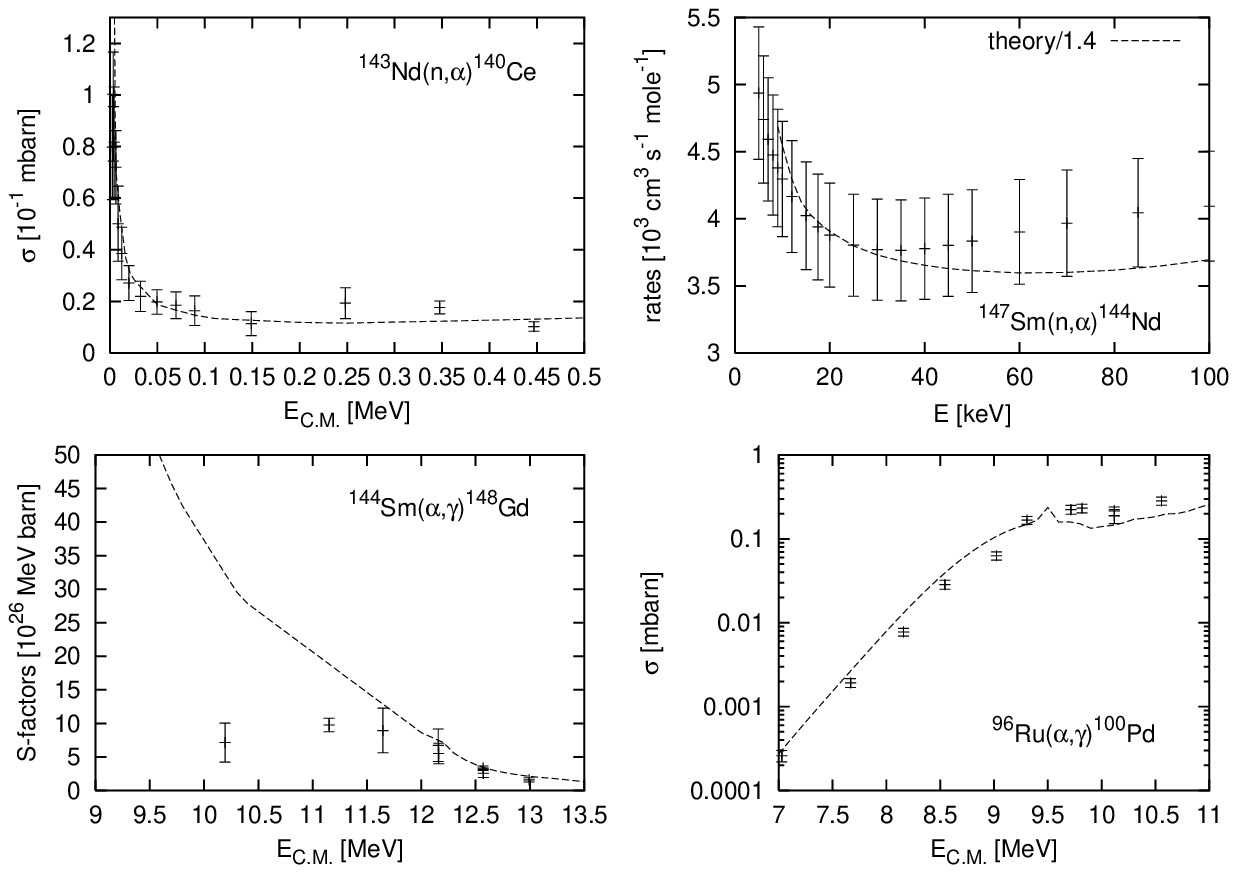}\par}

\caption{\label{figcarla3}Results for four different reactions using the new
energy-independent potential from \protect\cite{froh,froh1}
(see text). The dashed lines are the statistical model calculation. Note that
the \protect\( ^{147}\protect \)Sm(n,\protect\( \alpha \protect \)) result
is renormalized by a factor 1/1.4.}
\end{figure}

Despite the fact that the considered targets are in the same mass region, the
derived parameters also describe acceptably well the reaction \( ^{96} \)Ru(\( \alpha  \),\( \gamma  \))\( ^{100} \)Pd~\cite{ru}.
However, it is remarkable that even better overall agreement with all four reactions
can be obtained when using a mass- and energy-independent potential of Saxon-Woods
form for the real and imaginary parts (see Fig.\ \ref{figcarla3}). The real parameters are
given by \( V_{0}=162.3 \) MeV, \( r_{r}=1.27 \) fm, \( a_{r}=0.48 \) fm, the
imaginary ones by \( W_{0}(E)=W_{V}=25 \) MeV, \( r_{V}=1.4 \) fm, \( a_{V}=0.52 \)
fm. Thus, the real part is identical to the potential by Somorjai \textit{et
al.}~\cite{rau:som98} but without energy dependence, whereas the imaginary part
is similar to the one used in McFadden \& Satchler~\cite{rau:som98}. Since the
McFadden \& Satchler parameters were derived from extensive elastic scattering
data it seems reasonable that they are applicable to a wider range of targets.
The Somorjai \textit{et al.} parameters were derived for one reaction only but
seem to work also for the nuclides investigated here. The new
energy-independent potential also describes well other reactions not
shown here, e.g.\ $^{70}$Ge($\alpha$,$\gamma$)$^{74}$Se.

Here, we do not show our results from fitting each reaction separately. Obviously,
potentials fitted to single reactions can describe those -- but only those --
even better. 

Closer examination of Fig.\ \ref{figcarla3} seems to suggest that
an additional energy-dependence has to be introduced
at very low \( \alpha  \)-energies. This can be seen mainly from the
$^{144}$Sm($\alpha$,$\gamma$) comparison. However, since so far 
comparisons to other reactions were successful and the potentials only
failed for this one case, it could also be that this is just a peculiar
reaction. One peculiarity is the $\alpha$-decay of the final nucleus
$^{148}$Gd. In fact, in the considered reactions, also $^{144}$Nd is an
$\alpha$-emitter. As can also be seen from Figs.\
\ref{figcarla1}--\ref{figcarla3}, the description of
$^{147}$Sm($\alpha$,n)$^{144}$Nd is always less good than of those
reactions without $\alpha$-unstable final nucleus. This is supported by
the fact that non-statistical effects have been found in this reaction
\cite{koehnon}. It seems that an additional effect has to be included
for $\alpha$-unstable nuclides, either by a modified phenomenological
potential or by explicitly accounting for the correction. In the case of
$^{144}$Sm($\alpha$,$\gamma$), the effect seems to start acting
at around 12 MeV, perhaps similar to an additional barrier.

\section{Conclusions}

Stellar models have now reached a stage where the often quoted
astrophysical accuracy of ``a factor of two'' is not sufficient anymore
in many cases.
This poses a special challenge for the experimentalist as well as the
theoretician. Global nuclear models have already been quite successful
in predicting nuclear rates, especially for neutron- and proton-induced
reactions.
Despite these considerable successes
close to and far off stability, the description of certain nuclear inputs,
such as optical \( \alpha  \)-potentials, still needs to be improved. It is
also still unclear whether nuclear properties far off stability can be predicted
with sufficiently high accuracy. Although future advances in microscopic theories
may alleviate that problem, experimental data is clearly needed. Rare Isotope
Accelerators will make it possible to study highly unstable nuclides but also
{}``classical{}'' nuclear physics experiments with stable or long-lived nuclei
are indispensable. They are not only required to study specific crucial reactions
in stellar evolution and nucleosynthesis but can also
provide the systematics for global descriptions
and shed light on the interaction of different reaction mechanisms.

\section*{Acknowledgements}
I thank my collaborators Y. Alhassid, C. Fr\"ohlich, A. Heger, R. D.
Hoffman, G. Martinez-Pinedo, F.-K. Thielemann, and S. E. Woosley.
This work was supported in part by the Swiss NSF grants
2000-061031.02, 2024-067428.01.
I am also grateful for having received a PROFIL professorship from the 
Swiss NSF.

\end{document}